\documentclass[twocolumn,aps,showpacs]{revtex4-1}
\usepackage{amsfonts}
\usepackage{amsmath}
\usepackage{graphicx}
\usepackage{bm}
\usepackage{amssymb}
\usepackage{dcolumn}
\usepackage{color}

\begin{document}

\title{Surface-assisted Spin Hall Effect in Au Films with Pt Impurities}

\author{B. Gu$^{1,2}$, 
I. Sugai$^{3}$, 
T. Ziman$^{4}$, G. Y. Guo$^{5,6}$,
N. Nagaosa$^{7,8}$, 
T. Seki$^3$, K. Takanashi$^3$,
and S. Maekawa$^{1,2}$}

\affiliation{$^1$Advanced Science Research Center, Japan Atomic Energy Agency, Tokai 319-1195, Japan
\\ $^{2}$JST, CREST, 3-Sanbancho,Chiyoda-ku, Tokyo 102-0075, Japan
\\ $^{3}$Institute for Materials Research, Tohoku University, Sendai 980-8577, Japan
\\ $^{4}$CNRS and Institut Laue Langevin, Bo\^\i te Postale 156, F-38042 Grenoble Cedex 9, France
\\ $^{5}$Graduate Institute of Applied Physics, National Chengchi University, Taipei 116, Taiwan
\\ $^{6}$Department of Physics, National Taiwan University, Taipei 106, Taiwan
\\ $^{7}$Department of Applied Physics, The University of Tokyo, Tokyo 113-8656, Japan
\\ $^{8}$Cross-Correlated Material Research Group and Correlated Electron Research Group, RIKEN-ASI, Wako 315-0198, Japan}

\begin{abstract}
We show, both experimentally and theoretically, 
a novel route to obtain giant room temperature spin Hall effect due to 
surface-assisted skew scattering.
In the experiment, we report the spin Hall effect in Pt-doped
Au films with different thicknesses $t_N$.
The giant spin Hall angle 
$\gamma_S$ = $0.12 \pm 0.04$ is obtained for $t_N$ = 10 nm at room
temperature, while it is much smaller for  $t_N$ = 20 nm sample.
Combined {\it ab initio} and quantum Monte Carlo calculations 
for the skew scattering due to a Pt impurity show 
$\gamma_S$ $\cong$ 0.1 on the Au (111) surface,
while it is small in bulk Au.
The quantum Monte Carlo results show that
the spin-orbit interaction of the Pt impurity 
on the Au (111) surface is enhanced, 
because the Pt 5$d$ levels are lifted to the 
Fermi level due to the valence fluctuation.
In addition, there are two spin-orbit interaction channels on the Au (111) surface, 
while only one in bulk Au.
\end{abstract}

\pacs{71.70.Ej, 75.30.Kz, 75.40.Mg} \maketitle

The spin Hall effect (SHE) \cite{SHE-Hirsch}, 
which converts charge current into spin current in non-magnetic materials,  
is one of the key phenomena for the further development of spintronic devices.
From the viewpoint of practical applications,  materials are needed with a large spin Hall angle (SHA), 
the ratio between the induced spin Hall current and the input charge current. 
A recent experiment employing an Au Hall cross with an FePt perpendicular spin injector 
indicated a giant SHA of $\sim 0.1$ at room temperature \cite{AuFe-Takanashi}. 
On the other hand, quite a small SHA in Au was reported using a 60 nm thick 
Au Hall bar \cite{AuFe-Hoffmann1}. 

A possible mechanism of SHE in Au is the impurity scattering of electrons 
\cite{AuFe-GMN, AuCN-Mertig,AuFe-Gu}. 
In fact, Fert et al., have pointed out the importance of the skew scattering 
to both anomalous and spin Hall effects \cite{Au-Fert}.
In particular, the giant SHE is theoretically explained 
by the resonant skew scattering, i.e., spin-dependent deflection of the scattered electrons 
due to the spin-orbit interaction (SOI) of the Fe impurities in Au metal \cite{AuFe-GMN,AuFe-Gu}.
Experimentally, the effect of Fe doping on the SHE in Au was also investigated \cite{AuFe-Sugai}:
the SHA is approximately 0.07, and independent of the Fe concentrations,
all of which is in good agreement with the theories \cite{AuFe-GMN,AuFe-Gu}.

A previous paper \cite{AuFe-Seki} reported that SHA in undoped Au strongly 
depends on the thickness of the Au Hall cross, where Pt was not intentionally doped in Au.
This implies the importance of the surface and/or interface scattering,
because the thinner the film, 
the more efficient the surface scattering is.

In this Letter, we carry out a combined experimental and 
theoretical study on SHE in  Au films of two different thicknesses with
intentionally doped Pt impurities. 
We find the vital role of surface on the SHE,
which offers a new route to produce a  large SHE at room temperature. 
It is the $3rd$ route, in addition to the two known ones,
to give us a large SHE due to  skew scattering by impurities:
the $1st$ one originated from the simple and large SOI of impurities \cite{AHE-RMP}, 
and the $2nd$ one was rooted in the quantum renormalization by the  
Coulomb correlation $U$ or spin fluctuation of impurities in the bulk\cite{AuFe-GMN,AuFe-Gu}.

\begin{figure}[tbp]
\includegraphics[width = 8.5 cm]{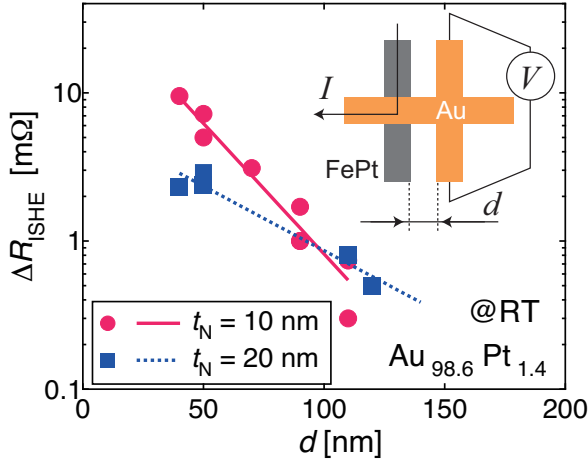}
\caption{ (color online) The resistance change of the inverse spin-Hall effect ($\triangle R_{ISHE}$)
as a function of the distance ($d$) between the Hall cross and the spin-injector. 
The Hall cross is composed of Pt-doped Au, and the concentration of Pt is 1.4 at $\%$. 
The thicknesses of the Hall crosses ($t_N$) are 10 nm (circles) and 20 nm (squares). 
The measurement was performed at room temperature. 
The solid and dotted lines are the results of the fitting for 
$t_N$ = 10 nm and 20 nm, respectively. 
The schematic illustration of the multi-terminal device is also shown in 
the inset of the figure. }
\label{F-Exp}
\end{figure}

\emph{Experimental giant SHE in Au films with Pt impurities.}---The thickness dependence of 
the SHA was investigated by measuring the inverse SHE in the lateral 
multi-terminal devices with a Pt-doped Au Hall cross. 
A schematic illustration of the device is shown in the inset of Fig. \ref{F-Exp}.  
The devices, consisting of an FePt perpendicular spin injector and a Pt-doped Au Hall cross 
with a predominantly (111) surface, were prepared on MgO (001) substrate. First, a 10 nm thick FePt 
layer showing perpendicular magnetization and a 10 nm thick Pt-doped Au layer were deposited on the 
substrate using an ultrahigh-vacuum magnetron sputtering system. A Pt-doped Au layer was prepared by 
the co-deposition from the Pt and Au targets. The shape of the FePt spin injector was patterned by 
electron beam lithography and Ar ion milling. Subsequently, a Pt-doped Au layer was again deposited 
on the patterned sample. Finally, the Pt-doped Au layer was patterned into a Hall cross. The widths 
of the spin injector and the Hall cross are 200 nm and 110 nm, respectively. A $dc$ electrical current 
was applied between the FePt spin polarizer and the lead of the Pt-doped Au wire, resulting in the 
pure spin current in the Pt-doped Au, and the voltage induced by SHE was measured using the Hall cross. 
(See Ref. \cite{AuFe-Takanashi} for more details).
The concentration of Pt in the Au Hall cross is 1.4 at$\%$,
which was determined by inductively coupled plasma mass spectrometry.
Figure \ref{F-Exp} shows the resistance change of the inverse SHE ($\triangle R_{ISHE}$) as a 
function of the distance ($d$) between the Hall cross and the spin-injector.
For both devices with the thicknesses ($t_N$) of 10 nm and 20 nm, 
$\triangle R_{ISHE}$ decreases exponentially as $d$ increases. 
By fitting the experimental data to the formula of
$ \triangle R_{ISHE} = (2\gamma_S\rho P/t_N)exp(-d/\lambda_N)$ (in Ref. \cite{AuFe-Takanashi}),
where $\rho$, $P$, $\lambda_N$ and $\gamma_S$ are 
the resistivity, the current spin polarization, the spin diffusion length, and SHA, respectively, 
$P$, $\lambda_N$ and $\gamma_S$ are estimated to be as follows: 
For $t_N$ = 10 nm, $P$ = 0.029, $\lambda_N$ = 25 $\pm$ 3 nm, and $\gamma_S$ = 0.12 $\pm$ 0.04. 
For $t_N$ = 20 nm, $P$ = 0.033, $\lambda_N$ = 50 $\pm$ 8 nm, and $\gamma_S$ = 0.008 $\pm$ 0.002. 
The values of $\rho$ were determined to be 
6.9 $\mu\Omega$cm and 6.0 $\mu\Omega$cm for $t_N$ = 10 nm and 20 nm, respectively, by measuring
the resistivity of the thin films prepared separately. 
It is noted that the large $\gamma_S$ is obtained for $t_N$ = 10 nm, 
which is larger than that obtained for undoped Au ($\gamma_S$ = 0.07) \cite{AuFe-Sugai}.
The increased $\rho$ with decreased $t_N$ shows the importance of the surface 
scattering \cite{SS-Maekawa}.
It is also clear that $\lambda_N$ decreases and $\gamma_S$ increases remarkably 
with decreased $t_N$.

\emph{Theoretical approach of SHE.}--- Because the resistivity of the 
sample is low ($\sim$ 5 $\mu\Omega$cm),
and the observed spin Hall conductivity is very large ($\sim$ 10$^4$ $\Omega^{-1}$cm$^{-1}$),
it is expected that the dominant contribution is due to the skew scattering,
and the side-jump contribution is small, 
as has been discussed for the anomalous Hall effect \cite{AHE-RMP}. 
From {\it ab initio} calculations \cite{Guo-JAP}, the intrinsic SHE is at least two orders of
magnitude smaller than the value observed here.

Here, by a combined theoretical approach \cite{JPCS-Gu},
we study the SHE due to the
skew scattering by a single Pt impurity both in bulk Au and
on Au (111) surface.
First, a single-impurity multi-orbital Anderson model
\cite{Anderson} is formulated within the 
density functional theory/local density approximation (DFT/LDA)\cite{DFT1,DFT2}, 
for determining the detailed host band structure, the
impurity levels, and the impurity-host hybridization. 
Second, the electron correlations in this Anderson model at finite
temperatures are calculated by the Hirsch-Fye quantum 
Monte Carlo (QMC) method\cite{QMC}.
The single-impurity multi-orbital Anderson model is defined as
\begin{eqnarray}
  H&=&\sum_{\textbf{k},\alpha,\sigma}\epsilon_{\alpha}(\textbf{k})
  c^{\dag}_{\textbf{k}\alpha\sigma}c_{\textbf{k}\alpha\sigma}
   +\sum_{\textbf{k},\alpha,\xi,\sigma}(V_{\xi\textbf{k}\alpha }
    d^{\dag}_{\xi\sigma} c_{\textbf{k}\alpha\sigma} + H.c.) \notag\\
  &+& \sum_{\xi,\sigma}\epsilon_{\xi}n_{\xi\sigma} 
   +U\sum_{\xi}n_{\xi\uparrow}n_{\xi\downarrow} 
   + \frac{U^{\prime}}{2}\sum_{\xi\neq\xi',\sigma,\sigma^{\prime}}
     n_{\xi\sigma}n_{\xi'\sigma^{\prime}} \notag\\
   &-& \frac{J}{2}\sum_{\xi\neq\xi',\sigma}n_{\xi\sigma}n_{\xi'\sigma}
  +\frac{\lambda}{2}\sum_{\xi,\sigma}d^{\dagger}_{\xi\sigma}
   (\ell)^{z}_{\xi\xi}(\sigma)^{z}_{\sigma\sigma}d_{\xi\sigma},
\label{E-Ham}
\end{eqnarray}
where 
$\epsilon_{\alpha}(\textbf{k})$ is host energy band,
$\epsilon_{\xi}$ is impurity energy levels, 
$V_{\xi\textbf{k}\alpha}$ is impurity-host
hybridization,
$U$ ($U'$) is the on-site Coulomb repulsion within (between) the orbitals of the
impurity, $J$ is the Hund coupling between the orbitals of the impurity,
and the last term is SOI, where for simplicity we consider only the $z$ component.
(See Ref.\cite{JPCS-Gu} for details). 

\begin{figure}[tbp]
\includegraphics[width = 8.5 cm]{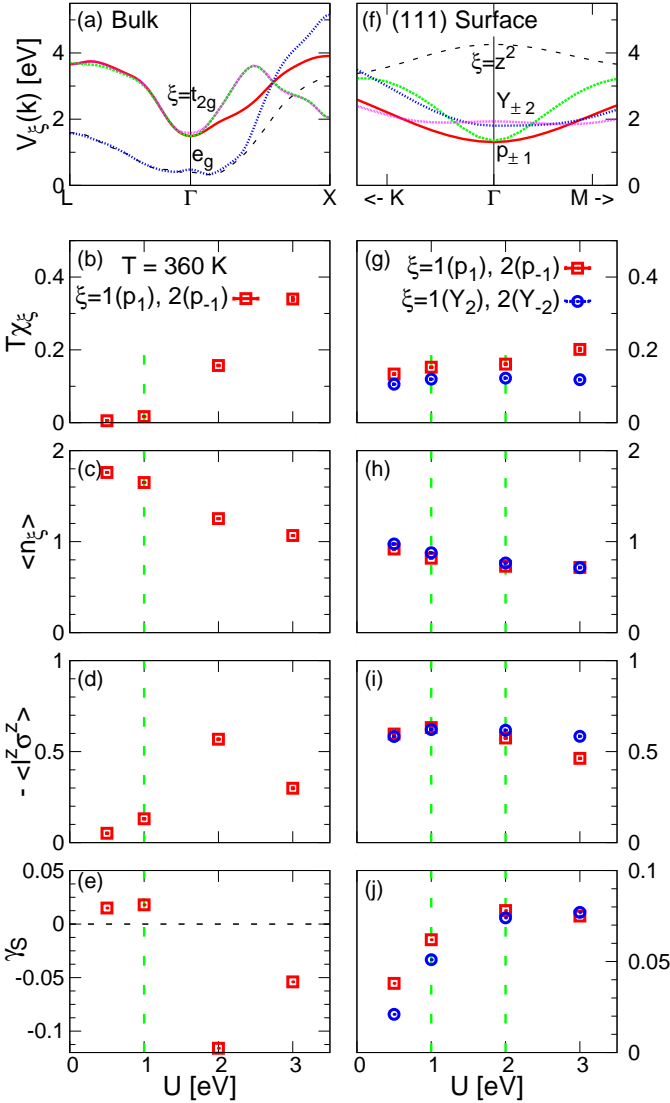}
\caption{ (color online) (a)-(e) are for a single Pt impurity doped in bulk Au.
(a) The hybridization function
between the $\xi$ orbitals of a Pt impurity and Au host,
obtained in the LDA calculations.
(b) The temperature times susceptibility $T\chi_{\xi}$,
(c) the occupation number $\langle n_{\xi}\rangle$,
and (d) the spin-orbit correlation function $-\langle \ell^z\sigma^z\rangle$
of the $\xi$ orbitals of a Pt impurity,
obtained by the QMC calculations at temperature $T$ = $360 K$.
(e) The calculated SHA $\gamma_S$. 
(f)-(j) are the counterparts of (a)-(e), respectively, for a single Pt impurity
doped on an Au (111) surface. The reasonable range of $U$ is between 1 and 2 eV, 
shown by vertical dashed lines. See text for details.}
\label{F-LDAQMC}
\end{figure}

\emph{A single Pt impurity in bulk Au.}---The LDA calculations are done by the
code {\sc Quantum-ESPRESSO} \cite{ESPRESSO}.
To obtain the hybridization of a Pt impurity in bulk Au,
we consider the supercell Au$_{26}$Pt, where a Pt atom 
is placed at the center of the supercell. 
(See Ref. {\cite{JPCS-Gu}} for details).
Fig. \ref{F-LDAQMC}(a) shows the hybridization function 
$V_{\xi}\left(\textbf{k}\right)$$\equiv$$\left(\sum_{\alpha}|V_{\xi\textbf{k}\alpha}|^2\right)^{1/2}$
between $\xi$ orbitals of a Pt impurity and bulk Au.
It is observed that, at the $\Gamma$ point ($\textbf{k}$=0), the
hybridization value of $\xi$ = $e_g$ ($z^2$,$x^2-y^2$) orbitals 
is much smaller than that of $\xi$ = $t_{2g}$ ($xz$,$yz$,$xy$) orbitals.
In the same LDA calculation, 
we have $\epsilon_{\xi}\cong$ -2.4 eV for $\xi$ = $e_g$, 
and $\epsilon_{\xi}\cong$ -2.3 eV for $\xi$ = $t_{2g}$
with zero Fermi energy.

A single Pt impurity has five $5d$ orbitals. 
Owing to the constraints of QMC calculations, 
we simplify it to a two-orbital model to
capture the essential physics.  
We consider the SOI within 
$p_1$ and $p_{-1}$ orbitals, 
where the notation corresponds to the
transformational properties of $t_{2g}$
orbitals equivalent to effective $p$ orbitals \cite{T2gP}: 
$p_1\equiv -\frac{1}{\sqrt{2}}(xz-iyz)$, $p_0\equiv -ixy$ and $p_{-1}\equiv
-\frac{1}{\sqrt{2}}(xz+iyz)$.
We do not consider the SOI within $x^2-y^2$ and $xy$ orbitals
since they are not degenerate \cite{AuFe-Gu}.
The last term of Eq.~(\ref{E-Ham})
is then written as $H_{so}=\frac{\lambda}{2}\ell^z\sigma^z$, 
where $\ell^z\sigma^z\equiv n_{1\uparrow}-n_{1\downarrow}
-n_{2\uparrow}+n_{2\downarrow}$,
and $\xi$ = 1(2) notes $p_1$($p_{-1}$) orbital.
The value of the SOI of $5d$ orbitals in a Pt atom
is  $\lambda$ = 0.4 eV \cite{AuPt-SOI}. 

To obtain the on-site Coulomb interaction parameter $U$
for Pt impurities in bulk Au, we do the QMC calculations with various $U$. 
According to the QMC calculations, 
the non-magnetic state, which is generally believed for 
Pt impurities in bulk Au \cite{AuPt-Sawatzky}, 
results in $U$ up to 1 eV \cite{AuPt-U},
as noted by a vertical dashed line in Figs. \ref{F-LDAQMC}(b)-(e).

For the $\xi$ orbitals of a Pt impurity doped in bulk Au,
Figs \ref{F-LDAQMC}(b)-(d) show the QMC results, 
at temperature $T$ = $360 K$, 
of the temperature times susceptibility $T\chi_{\xi}$
with  $\chi_{\xi} \equiv \int_{0}^{\beta}d\tau \langle
M^z_{\xi}(\tau)M^z_{\xi}(0)\rangle$ and
$M^z_{\xi} \equiv n_{\xi\uparrow} - n_{\xi\downarrow}$,
the occupation number $\langle n_{\xi}\rangle$
with $n_{\xi} \equiv n_{\xi\uparrow} + n_{\xi\downarrow}$,
and the spin-orbit correlation function $-\langle \ell^z\sigma^z\rangle$
as defined above. 
Based on these QMC estimates,
we calculate the $\gamma_S$ as in Ref. \cite{AuFe-Gu}.
As shown in Figs. \ref{F-LDAQMC}(c)-(e), for $U$ = 1 eV it has 
$n_1$ = $n_2$ = 1.65,
$\langle \ell^z\sigma^z\rangle$ = -0.13,
and $\gamma_S$ = 0.018. 

\emph{A single Pt impurity on an Au (111) surface.}---To calculate the hybridization 
of a Pt impurity on an Au (111) surface,
we consider the supercell Au$_{71}$Pt, which consists of 24 layers
with 3 atoms per layer($\sqrt{3}\times\sqrt{3}R30^{\circ}$), and a Pt atom is 
placed at the center of the top layer. 
Fig. \ref{F-LDAQMC}(f) shows the hybridization between $\xi$ orbitals
of a Pt impurity and the Au (111) surface.
In the following discussion we shall take  $x$ and $y$ axes in the Au (111) surface, 
and $z$ axis to be  normal. 
We note that at the $\Gamma$ point ($\textbf{k}$=0), 
the hybridizations of $\xi$ = $x^2-y^2$ and $xy$ orbitals of Pt are 
nearly the same [$Y_2\equiv \frac{1}{\sqrt{2}}(x^2-y^2+ixy)$
and $Y_{-2}\equiv \frac{1}{\sqrt{2}}(x^2-y^2-ixy)$],
in contrast to the bulk case shown in Fig. \ref{F-LDAQMC}(a). 
Indeed, in the same LDA calculations
we obtain the nearly degenerate orbitals $\xi$ = $xz$ and $yz$ with
$\epsilon_{\xi}\cong$  -0.3 eV, and another pair of nearly degenerate 
orbitals $\xi$ = $x^2-y^2$ and $xy$ with $\epsilon_{\xi}\cong$ -0.2 eV. 

The nearly degenerate $xz$ and $yz$ orbitals 
give the SOI channel of $p_1$ and $p_{-1}$ with $\ell^z$ = $\pm1$,
and the nearly degenerate $x^2-y^2$ and $xy$ orbitals 
give another SOI channel of $Y_2$ and $Y_{-2}$ with $\ell^z$ = $\pm2$.
Owing to the constraints of QMC calculations,
we use the two-orbital model to 
study the SOI channels of
$p_{\pm 1}$ and $Y_{\pm 2}$, respectively. 
For $\xi$ = 1(2) notes $p_1$($p_{-1}$) orbital,
the last term of Eq.~(\ref{E-Ham})
is written as $H_{so}=\frac{\lambda}{2}\ell^z\sigma^z$.
For $\xi$ = 1(2) notes $Y_2$($Y_{-2}$) orbital,
it is written as $H_{so}=\lambda\ell^z\sigma^z$.

Figs \ref{F-LDAQMC}(g)-(i) show the QMC results at $T$ = 360 K
of the $T\chi_{\xi}$,  
$\langle n_{\xi}\rangle$ and 
$-\langle \ell^z\sigma^z\rangle$
of the $\xi$ orbitals of a Pt impurity doped on the Au (111) surface. 
In bulk Au,
the reasonable parameter $U$ of Pt impurities is $\sim$ 1 eV.
On the surface, the $U$ of Pt impurities could increase 
because of the decreased screening effect there. 
Thus, as noted by vertical dash lines in Figs. \ref{F-LDAQMC}(g)-(j),
the reasonable range of $U$ for Pt impurities on Au surface may be 1 $\sim$ 2 eV.

We now calculate $\gamma_S$
for channels of $p_{\pm 1}$ and $Y_{\pm 2}$, respectively,
as in Ref. \cite{AuFe-Gu}.
As shown in Figs. \ref{F-LDAQMC}(h)-(j), 
for $\xi$ = $p_{\pm 1}$ orbitals and $U$ = 1 eV, it has
$n_1$ = $n_2$ = 0.82, $\langle \ell^z\sigma^z\rangle$ = -0.63, 
and $\gamma_S$ = 0.062;
for $\xi$ = $Y_{\pm 2}$ orbitals and $U$ = 1 eV, it has
$n_1$ = $n_2$ = 0.87, $\langle \ell^z\sigma^z\rangle$ = -0.62, 
and $\gamma_S$ = 0.051.
For a larger parameter $U$ = 2 eV, 
a larger SHA is obtained as 
$\gamma_S$ = 0.078 for $\xi$ = $p_{\pm 1}$ orbitals
and 
$\gamma_S$ = 0.074 for $\xi$ = $Y_{\pm 2}$ orbitals.
As an approximation, the total $\gamma_S$ of the Pt impurity
on the Au (111) surface
could be estimated as the sum of that of $p_{\pm 1}$ and $Y_{\pm 2}$ channels.
We then have the total
$\gamma_S$ = 0.11 for $U$ = 1 eV 
and $\gamma_S$ = 0.15 for $U$ = 2 eV. 

The QMC results show that 
$n_{\xi}$ ($\xi$ = $p_{\pm 1}$, $Y_{\pm 2}$) 
of the Pt impurity decrease from $\sim 2$ in bulk Au 
to $\sim 1$ on the Au (111) surface,
implying that the $\xi$ levels of the Pt impurity on 
the Au (111) surface are lifted to the Fermi level 
due to the valence fluctuation. 
As a result, the SOI in two channels of $\ell^z$ = $\pm 1$ and 
$\ell^z$ = $\pm 2$ is enhanced, 
and the large SHE is obtained. 
This theoretically obtained $\gamma_S$ for a Pt impurity on the Au(111) surface 
is consistent with the magnitude and sign of the experimentally obtained $\gamma_S$. 
Note that predicted $\gamma_S$ due to a Pt impurity in bulk Au is also of the order 
of -0.1 for $U$ = 2 eV (Fig. 2(e)). 
However, we believe this value of $U$ is larger than that for 5$d$ orbital of Pt, 
and also the sign of $\gamma_S$ is the opposite to the experimental one.

\emph{Discussion.}---First, we discuss the relation between 
the $\lambda_N$ and $t_N$ observed in the experiment. 
It is known that 
$\lambda_N=\sqrt{D\tau_N}$, 
where $D$ is the diffusion constant and
$\tau_N$ is the spin flip relaxation time \cite{SM-Spin}. 
The golden rule gives
$1/\tau_N$$\propto$$\langle H_{so}\rangle^2$,
where $H_{so}$ is the SOI Hamiltonian due to the 
impurity scattering \cite{SM-Spin}.
Thus we have $\lambda_N\propto |\langle H_{so} \rangle|^{-1}$.
As suggested by the $d$-dependence in Fig.1, 
the surface scattering is more efficient for the thinner film, 
and hence the $|\langle H_{so}\rangle|$ is larger and $\lambda_N$ is shorter.

Next, we note that Rashba spin-orbit splitting on 
the Au (111) surface is known to be large \cite{Au111-Reinert}.
However, the spin Hall conductivity due to this Rashba interaction
\cite{Sinova} is an order of magnitude smaller than that observed here.
 Therefore, the Rashba interaction is not the main source of the 
giant SHA.
 
Third, we discuss  a single Pt impurity on an Au (001) surface. 
The QMC results show that
$n_{\xi}$ ($\xi$ = $p_{\pm 1}$) of the Pt impurity decrease from $\sim 2$ in bulk Au
to $\sim 1$ on the Au (001) surface, 
similarly to the case of Au(111) surface as discussed above.
Thus, the SOI in the channel of $\ell^z$ = $\pm 1$ is enhanced.
Because of the Au (001) surface symmetry, 
the $x^2-y^2$ and $xy$ orbitals of the Pt impurity are not degenerate, 
and there is no SOI channel of $\ell^z$ = $\pm 2$.
Accordingly, the spin Hall current and $\gamma_S$ on the Au (001) surface
are only {\it a half} of those on the Au (111) surface.   
This could be tested experimentally.

We note that we now have two routes leading to the giant 
SHE as originally observed in the ``undoped Au''\cite{AuFe-Takanashi,AuFe-Seki}:
the orbital-dependent Kondo effect on Fe impurities \cite{AuFe-GMN,AuFe-Gu}, 
and a new one: surface-assisted skew scattering on Pt impurities.
Probably both mechanisms contributed to the SHE in the original experiments on
``undoped'' samples, 
but the new mechanism would better explain the thickness dependence observed there \cite{AuFe-Seki}. 
On the other hand, the new samples of Pt-doped Au give unambiguous evidence for the 
new route. 

To conclude, we show, both experimentally and theoretically, 
a novel route to obtain giant room temperature SHE due to 
the surface-assisted skew scattering.
In the experiments, we report the SHE in Pt-doped
Au films with different $t_N$.
The giant SHA $\gamma_S$ = $0.12 \pm 0.04$ is obtained for $t_N$ = 10 nm at room
temperature, while it is much smaller for  $t_N$ = 20 nm sample.
In the combined {\it ab initio} and QMC calculations
for the skew scattering due to a Pt impurity,
we show that $\gamma_S$ $\cong 0.1$ on the Au (111) surface,
while it is small in bulk Au.
We find that 
(i) there are two SOI channels for Pt atom on the Au (111) surface,
while only one in bulk Au,
and (ii) the QMC results show that
$n_{\xi}$ ($\xi$ = $p_{\pm 1}$, $Y_{\pm 2}$)
of the Pt impurity decrease from $\sim 2$ in bulk Au
to $\sim$ 1 on the Au (111) surface,
implying that the $\xi$ levels of the Pt impurity on
the Au (111) surface are lifted to the Fermi level
due to the valence fluctuation.
As a result, the SOI in two channels of $\ell^z$ = $\pm 1$ and
$\ell^z$ = $\pm 2$ is enhanced. 
Combined (i) and (ii), 
the large SHE is obtained for the Pt impurity on the Au (111) surface.

This work was supported by Grant-in-Aids for Scientific Research in 
the priority area ``Spin current'' with No. 19048015, No. 19048009, No. 1904008, 
and No. 21244053,  
the Next Generation Super Computing Project, 
the Nanoscience program from MEXT of Japan, 
the Funding Program for World-Leading Innovative R$\&$D on 
Science and Technology (FIRST program),
and NSC and NCTS of Taiwan. 
Prof. N. Bulut prepared the original code of QMC program.  
We thank Dr. S. Mitani for his helpful comments.


\begin{thebibliography}{99}
\bibitem{AuPt-U} In addition, $U_{eff}\equiv U-J$= $0.7 eV$ ($U$= $1 eV$, $J$= $0.3 eV$) 
is in agreement with $U_{eff}$= $0.5 eV$ used in Ref. \cite{AuFe-GMN}.
\end{thebibliography}
\end{document}